\documentstyle{mn}

\newif\ifAMStwofonts

\ifoldfss
  \ifCUPmtlplainloaded \else
   \NewTextAlphabet{textbfit} {cmbxti10} {}
    \NewTextAlphabet{textbfss} {cmssbx10} {}
    \NewMathAlphabet{mathbfit} {cmbxti10} {} 
    \NewMathAlphabet{mathbfss} {cmssbx10} {} 
  \fi
  \ifAMStwofonts
    \ifCUPmtlplainloaded \else
      \NewSymbolFont{upmath} {eurm10}
      \NewSymbolFont{AMSa} {msam10}
      \NewMathSymbol{\upi}     {0}{upmath}{19}
      \NewMathSymbol{\umu}     {0}{upmath}{16}
      \NewMathSymbol{\upartial}{0}{upmath}{40}
      \NewMathSymbol{\leqslant}{3}{AMSa}{36}
      \NewMathSymbol{\geqslant}{3}{AMSa}{3E}

    \fi
  \fi
\fi 

\ifnfssone
  \newmathalphabet{\mathit}
  \addtoversion{normal}{\mathit}{cmr}{m}{it}
  \addtoversion{bold}{\mathit}{cmr}{bx}{it}
  \newmathalphabet{\mathbfit} 
  \addtoversion{normal}{\mathbfit}{cmr}{bx}{it}
  \addtoversion{bold}{\mathbfit}{cmr}{bx}{it}
  \newmathalphabet{\mathbfss} 
  \addtoversion{normal}{\mathbfss}{cmss}{bx}{n}
  \addtoversion{bold}{\mathbfss}{cmss}{bx}{n}
  \ifAMStwofonts
    \ifCUPmtlplainloaded \else
      \UseAMStwoboldmath
      \makeatletter
      \new@mathgroup\upmath@group
      \define@mathgroup\mv@normal\upmath@group{eur}{m}{n}
      \define@mathgroup\mv@bold\upmath@group{eur}{b}{n}
      \edef\UPM{\hexnumber\upmath@group}
      \new@mathgroup\amsa@group
      \define@mathgroup\mv@normal\amsa@group{msa}{m}{n}
      \define@mathgroup\mv@bold\amsa@group{msa}{m}{n}
      \edef\AMSa{\hexnumber\amsa@group}
      \makeatother
      \mathchardef\upi="0\UPM19
      \mathchardef\umu="0\UPM16
      \mathchardef\upartial="0\UPM40
      \mathchardef\leqslant="3\AMSa36
      \mathchardef\geqslant="3\AMSa3E
    \fi
  \fi
\fi 

\ifnfsstwo
  \DeclareMathAlphabet{\mathbfit}{OT1}{cmr}{bx}{it}
  \SetMathAlphabet\mathbfit{bold}{OT1}{cmr}{bx}{it}
  \DeclareMathAlphabet{\mathbfss}{OT1}{cmss}{bx}{n}
  \SetMathAlphabet\mathbfss{bold}{OT1}{cmss}{bx}{n}
  \ifAMStwofonts
    \ifCUPmtlplainloaded \else
      \DeclareSymbolFont{UPM}{U}{eur}{m}{n}
      \SetSymbolFont{UPM}{bold}{U}{eur}{b}{n}
      \DeclareSymbolFont{AMSa}{U}{msa}{m}{n}
      \DeclareMathSymbol{\upi}{0}{UPM}{"19}
      \DeclareMathSymbol{\umu}{0}{UPM}{"16}
      \DeclareMathSymbol{\upartial}{0}{UPM}{"40}
      \DeclareMathSymbol{\leqslant}{3}{AMSa}{"36}
      \DeclareMathSymbol{\geqslant}{3}{AMSa}{"3E}
    \fi
  \fi
\fi 

\ifCUPmtlplainloaded \else
  \ifAMStwofonts \else 
    \def\upi{\pi}
    \def\umu{\mu}
    \def\upartial{\partial}
  \fi
\fi

\title{A Phenomenological Model for the Evolution of Proto-Galaxies}
\author[F. Tabatabaei and S.Nasiri]
       {F.Tabatabaei$^1$
       \& S.Nasiri$^{1,2}$
       \\ $^1$Institute for Advanced Studies in Basic Sciences, Zanjan, Iran\\
       $^2$Department of Physics, Zanjan University, Zanjan, Iran}
\date{}

\pagerange{\pageref{firstpage}--\pageref{lastpage}}
\pubyear{}

\begin{document}

\maketitle

\label{firstpage}








\begin{abstract}

The contraction model of Field and Colgate for proto-galaxies, first proposed
to describe the observed properties of quasars, is generalized and used to investigate
the evolution of galaxies. The LEDA data base for elliptical, spiral, compact
and diffuse galaxies is employed and it is shown that the above model is consistent with
observational evidences regarding their dynamical evolution, star formation
rate and different morphologies.

\end{abstract}


\begin{keywords}
galaxies: formation -- galaxies: luminosity function -- quasars: general --galaxies:
phenomenology.
\end{keywords}

\section{INTRODUCTION}

It is generally thought that the galaxies are formed by cooling and condensation
from the intergalactic gas clouds of sufficiently high densities [1,2,3,4].
The recent great advances in observational technologies made the proto-galaxies (PGs) one of the
most exciting and lively fields of extragalactic astronomy. Rees and Ostriker [5]
argued that the collapse of a virialised gas for which the cooling time is shorter than dynamical time
will lead to the formation of a galaxy. White and Rees [6] solved the problem of
cooling catastrophe hierarchical model of Rees and Ostriker, by introducing the idea
of feedback resulting from energy release from supernovae associated with the early
generation of stars and reheating the gas before having a chance to condense.
On the other hand, the initial total mass [7], the initial angular momentum [8]
and the initial density distribution [9] play a significant role in evolution
history of the galaxies. Here, we consider the Field and Colgate (FC) model [10] and generalized
it to include the effect of initial mass and initial size as well as initial angular velocity.
To do this, and to have a conserved dynamical quantity during the collapse
of the cloud, we introduce the specific angular momentum (SAM) as
parameter which governs the evolution. Furthermore, it is known that the star formation history of the galaxies depends
on their rotations [11]. Therefore, one expects that the observed luminosities of the
galaxies would be affected by star formation rate (SFR), which itself may depend on
SAM. From this point of view, SFR may be included in the FC model. The generalized  version
of the FC model (in the form mentioned above), called GFC model, is used to explain
the observed properties of galaxies such as, their morphologies, luminosities,
compactness, evolution and SFR. The LEDA database
of 100,000 galaxies [12] is used as our source data. It is shown that the SAM of PGs has a significant
effect on their evolution. In addition, the different values of SAM for different
morphologies, together with their SFR may be employed to present a conjecture
about their origin.
\\
In section 2 the FC model is reviewed and is generalized to introduce the GFC model.
In section 3 the GFC model is examined versus the observational data. It is shown that
this model is capable of explaining various aspects of galaxies. Section 4
is devoted to the concluding remarks.

\section{GFC MODEL}

The FC model was first introduced to describe the observational properties of
quasars. This model assumes that for proto-galaxies of the same mass and size, the
average mass of their constituent stars, their total energy output and their
final size at the end of galaxy formation depend on their initial angular
velocities. According to this model, the proto-galaxies with lower
angular velocities would be eventually converted to more compact and
luminous objects such as quasars and compact galaxies. This proposition may
be tested by using the observational data obtained in galaxy surveys. However, before
doing this, a few points must be taken into account. In contrast to
the assumptions of FC model, the angular velocity is not a conserved
quantity during the contraction processes. Therefore, the observed angular
velocities do not indicate any relevance to their initial values which the
proto-galaxy starts the contraction with. Thus another parameter must be
replaced by angular velocity which keeps a constant value during the
contraction. Another point is that the typical mass and size assumed for all
proto-galaxies in FC model do not make any sense. In the other words, one
actually expects some kind of distribution for these quantities rather than
a fixed value. Thus upon these considerations, we propose the angular
momentum (i.e. angular momentum per unit mass) instead of angular velocity
and call the revised version of FC model as generalized FC or GFC model. However,
the specific angular momentum (hereafter SAM) is assumed to be
constant during the contraction process. This assumption does not essentially
change the dynamical grounds of the FC model and revise it only to obtain a better fit
to observations. In fact, FC model will emerge as a special case
from GFC model. In the framework of GFC model, a given SAM may correspond
to different values for masses and sizes of proto-galaxies. Now the lower
(higher) values of SAM will eventually lead to more compact (diffuse) and
luminous (faint) galaxies. These will be confirmed observationally in the
next sections. As mentioned before, the FC model is proposed to describe the
observational properties of quasars. However, because of the lack of required
data for these objects, one may use the already available data for galaxies. If
the model is shown to work well for galaxies, one might try to extend it for
quasars, too. On the other hand, compact objects will possess much more
massive stars. Therefore, one should argue that by rapid energy consumption
they must evolve faster than those having low mass stars. This processes if done for
quasars which by GFC model are extreme case of galaxies with less SAM,
will eventually make them to disappear as a result of the collapse followed by
consuming their energy sources. Thus, GFC model predicts an evolutionary
''decay mechanism'' for quasars in the course of time. We will show that
this process may be equally applied for compact galaxies and those with
lower SAM, which, in turn, have young blue stars and less
interstellar gas.

\section{GFC MODEL VERSUS OBSERVATION}

Let us investigate the dependence of SAM for different morphologies and
types of galaxies on their SL. The data is extracted from LEDA data base.
First, we compare the behavior of SAM versus SL for the ellipticals and spirals
as two distinct morphologies. Then the same thing is done for different
types of spirals, according to the de Vaucouleurs classification.

\subsection{GALAXIES WITHIN THE SAME MORPHOLOGICAL TYPE}

According to the GFC model, we expect that the luminosities of galaxies increase as
their SAM decrease. Figure 1 shows the behavior of SAM for 56 elliptical
galaxies in terms of their luminosity. The result shown in Fig.~1, however, is not consistent
with GFC model.

To get the expected result one might
use the specific luminosity (luminosity per unit mass),
instead of luminosity itself for the galaxies. The results obtained in such
a way are shown in Fig.~2.

 The best fitted curve to this figure has the form
$y=a+b\exp (-{\frac xc})$, with $a=1.05$, $b=34.41$, $c=0.4$. Here, $y$ is specific
luminosity in units of $({\frac{L_{\odot }}{M_{\odot }}})$ and x is SAM in
units of $(pc^2yr^{-1})$.

\begin{figure}
 \vbox to2.5in{\rule{0pt}{2.5in}}
\includegraphics{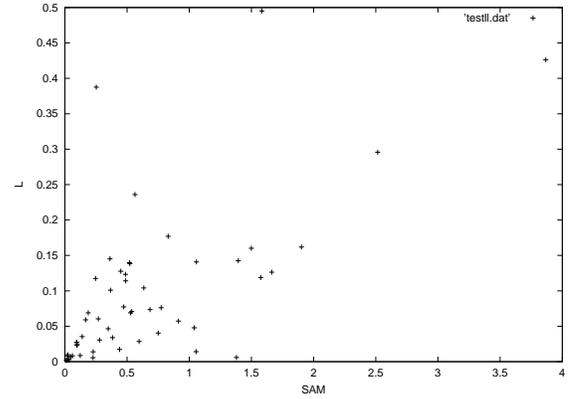}
\caption{ Variation of luminosity (L) of 56 elliptical galaxies (in units of $10^{10} L\odot$)
versus their SAM ($pc^{2}yr^{-1}$)}.
\end{figure}

\begin{figure}
\vbox to2.5in{\rule{0pt}{2.5in}}
\includegraphics{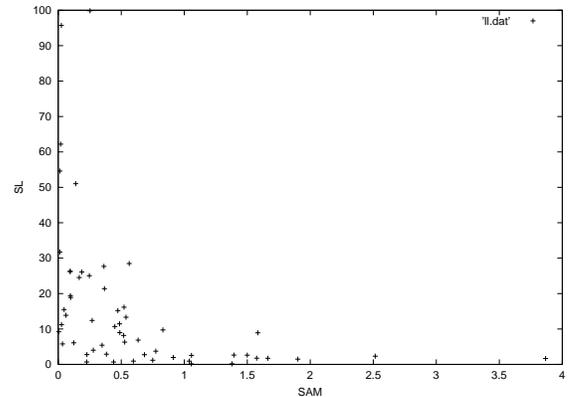}
\caption{Variation of SL (${L\odot\over M\odot}$) versus SAM for elliptical galaxies.}
\end{figure}

 The exponential behavior shows the decreasing
luminosity with increasing SAM, in agreement with the GFC hypothesis. The same
quantities are obtained for different morphologies of de Vaucouleurs class of
spiral galaxies and the results are plotted in Figs.~3 to ~6. Again, the
results support the GFC model. Of course,
the functional form of SAM are not the same for all morphologies. The best
fitted curve for 226 galaxies of Sa type (Fig.3) is, $y=a+b\exp (-x)$ where $%
a=3.43$, $b=46.04$, $c=0.76$ and for 717 galaxies of Sb type (Fig.4) is, $%
y=a+b\exp (-{\frac xc})$ where, $a=4.88$, $b=113.75$, $c=0.59$ and for 1543
galaxies of Sc type (Fig.5) is, $y=a+b\exp (-{\frac xc})$ where, $a=4.89$, $%
b=24.31$, $c=1.61$. For all of the above Figures, there are an exponential best
fitted curve showing a behavior similar to the one for Sa type. For 389
Sd type galaxies (Fig.6) the best fitted curve is in the form, $y=a+{\frac
b{(x^{{\frac 12}})}}$ where, $a=5.38$,$b=6.79$ and for 285 Sm type galaxies,
$y=a+{\frac bx}$ where $a=12.54$, $b=2.3$. These figures show a negative
power form for variation of SAM and SL. \\

\begin{figure}
\vbox to2.5in{\rule{0pt}{2.5in}}
\includegraphics{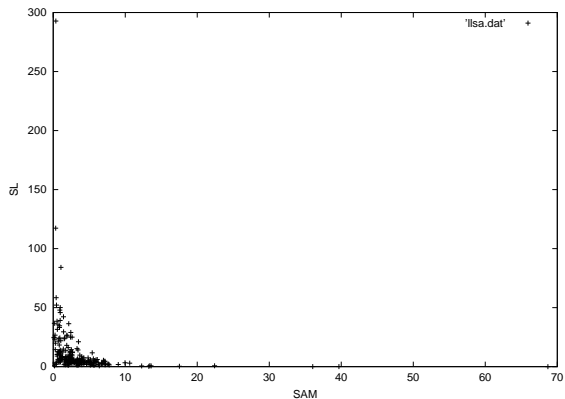}
    \caption{ The same as fig. ~2 for 226 Sa type galaxies.}
 \end{figure}

\begin{figure}
\vbox to2.5in{\rule{0pt}{2.5in}}
\includegraphics{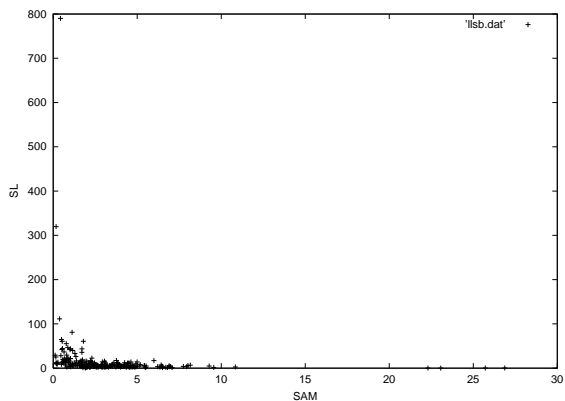}
 \caption{ The same as fig.~2 for 717 Sb type galaxies.}
\end{figure}

\begin{figure}
\vbox to2.5in{\rule{0pt}{2.5in}}
\includegraphics{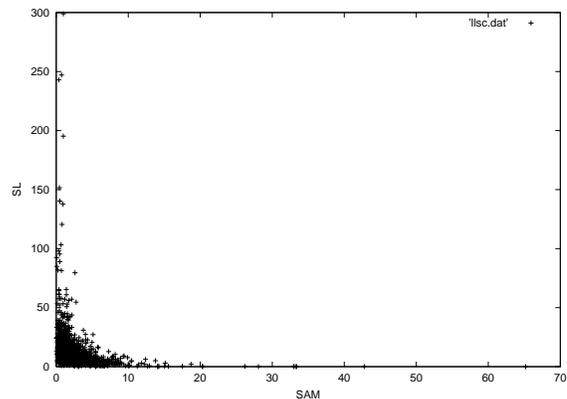}
\caption{The same as fig.~2 for 1534 Sc type galaxies.}
 \end{figure}

\begin{figure}
\vbox to2.5in{\rule{0pt}{2.5in}}
\includegraphics{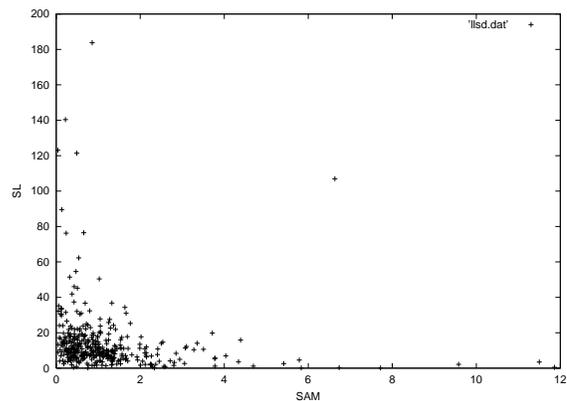}
 \caption{The same as fig. ~2 for 389 Sd type galaxies.}
\end{figure}

\subsection{ELLIPTICALS VERSUS SPIRALS}

Spiral galaxies rotate faster than elliptical galaxies[14].
According to GFC model their luminosity is expected to be less.
However, this is not supported by low redshift galaxy observations. The
discrepancy seems to be removed as follows. In general, most of the
observations are limited to the near distances or
low redshifts. However, the high redshift observations show that the
evolution procedure is not the same for spirals and ellipticals. In other
words, they are in different phases of star formation. Ellipticals have
stopped their star formation processes because of fast SFR in the past,
while spirals are still active and therefore, have some young blue stars
producing the observed excess luminosity. Therefore, the final elliptical or
luminous spirals not only violate the GFC model, but rather
confirm it. That is, ellipticals were started from slowly rotating
proto-galaxies compared with those of spirals. Therefore the condensation
rate for ellipticals were higher than spirals leading to higher SFR in a
certain phase of evolution for ellipticals. As a result, ellipticals evolved
faster than spirals and now contain old stars, showing in some sense, the
decay mechanism for galaxies. The same mechanism works for quasars with
higher decay rates and it will be investigated elsewhere[15].

\subsection{DE VAUCOULEURS TYPES OF SPIRALS}

We now look for a relation between SAM and
luminosity for different de Vaucouleurs types of spiral galaxies. In Figs.~7 to
~13 we illustrate the absolute magnitude or luminosity distribution of spiral
galaxies of different de Vaucouleurs type. We may fit Gaussian distributions to
these figures and calculate the most probable absolute magnitude, $M_{mp}$,
for which the number of galaxies is maximum, for each group belonging to
a specific morphology. Results for $M_{mp}$ as well as the average value
of SAM for each type are given in the table. It is seen that SAM increases
from Sbc to Sd. However, it fluctuates from Sa to Sbc, whereas the $M_{mp}$
decreases from Sb to Sd and fluctuates from Sa to Sb. If one resorts the
table, say by increasing order of SAM, one can find that the $L_{mp}$
increases with decreasing average SAM.

 On the other hand, the average mass, $
\overline{m}$, corresponding to each type given in the 4th column of the
table increases in the same manner as $M_{mp}$ from Sbc to Sd and
fluctuates from Sa to Sbc. This phenomenological investigation results shows
the role of mass on affecting the $M_{mp}$. While the ratio of ${\frac{%
L_{mp}}{\overline{m}}}$ ,shown in 5th column of the table, increases from
Sbc to Sd and fluctuates from Sa to Sbc. This apparently contradicts with
GFC model. However, similar to what we did in section 3.2 for comparison
of ellipticals and spirals, one can argue as follows:
It is known that the
late type spirals are those with greatest amount of dust and gas and highest
SFR. This gives the lowest ''mass-luminosity'' ratio for these type of
galaxies which in turn leads to the observed behavior. Thus one may conclude
that the SFR plays a significant role in evolution and distribution of
galaxies. In the other words, according to the GFC model we expect that
for those morphologies of spiral galaxies with less SAM, undergo more rapid
"decay mechanism" compared with those having higher SAM. This leads to lower present
SFR giving lower SL. Therefore, observed properties not only do not contradict
with the GFC model, but also support it.
\\
Also the effect of SAM on SFR discussed above is in agreement with that of
Samland and Hensler [12] obtained by a chemo-dynamical approach.

\subsection{COMPACTNESS AND GFC MODEL}

We finally use GFC model to study the observational properties of compact (C)
and diffuse (D) galaxies given by LEDA data base. Figs.~14 and ~15 show the distribution
of absolute magnitude for 450 C-type and 230 D-type galaxies, respectively.
It is seen that the most probable magnitudes are $\sim$-20.82 and -19.76 for
C- and D-type galaxies, respectively. Therefore, the compact galaxies are on the average
about one magnitude more luminous than diffuse galaxies, suggesting another
observational evidence satisfied by GFC model.

\begin{figure}
\vbox to2.5in{\rule{0pt}{2.5in}}
\includegraphics{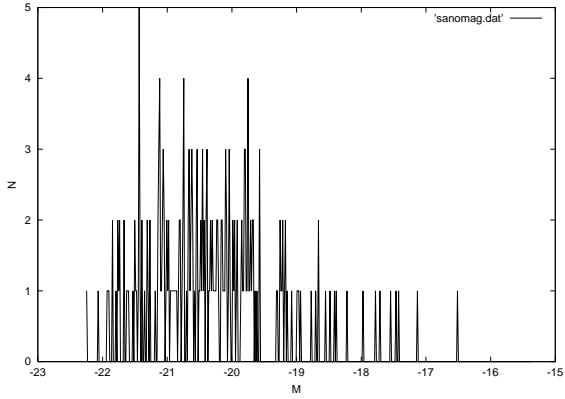}
\caption{Distribution of the Sa type galaxies in terms of their absolute magnitude.}
\end{figure}

\begin{figure}
\vbox to2.5in{\rule{0pt}{2.5in}}
\includegraphics{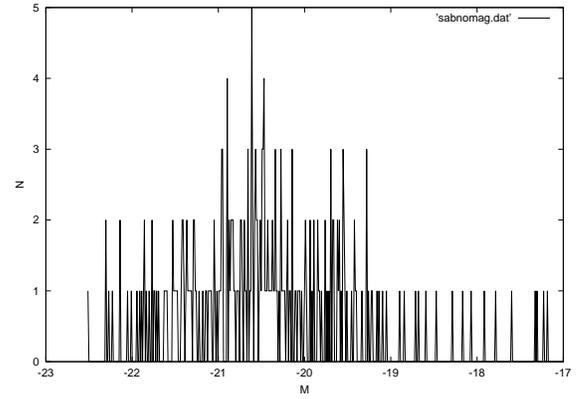}
\caption{ The same as fig.~7 for about 248 Sab galaxies.}
\end{figure}

\begin{figure}
\vbox to2.5in{\rule{0pt}{2.5in}}
\includegraphics{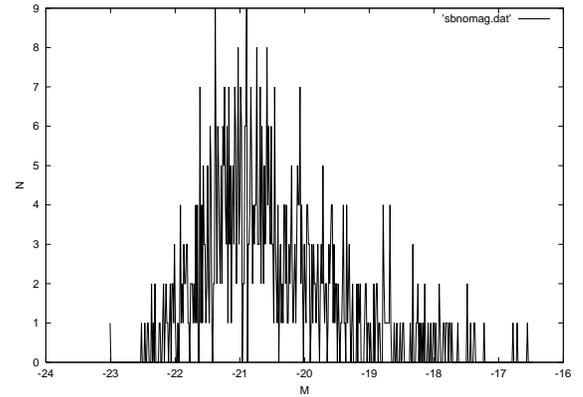}
\caption{The same as fig.~7 for about 717 the Sb type galaxies.}
\end{figure}

\begin{figure}
\vbox to2.5in{\rule{0pt}{2.5in}}
\includegraphics{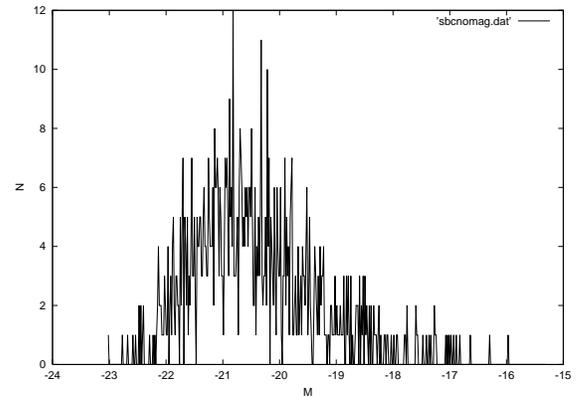}
\caption{The same as fig.~7 for about 824 Sbc type galaxies.}
\end{figure}

\begin{figure}
\vbox to2.5in{\rule{0pt}{2.5in}}
\includegraphics{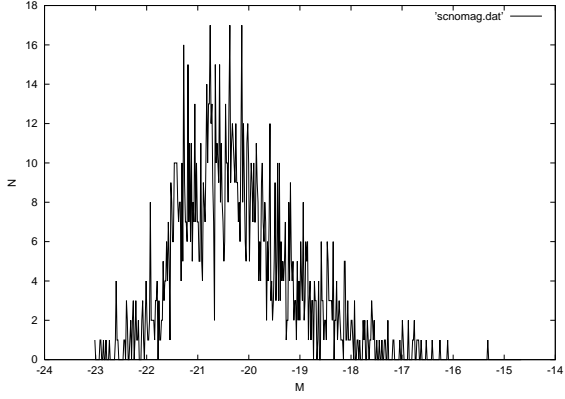}
\caption{The same as fig.~7 for about 1534 Sc type galaxies.}
\end{figure}

\begin{figure}
\vbox to2.5in{\rule{0pt}{2.5in}}
\includegraphics{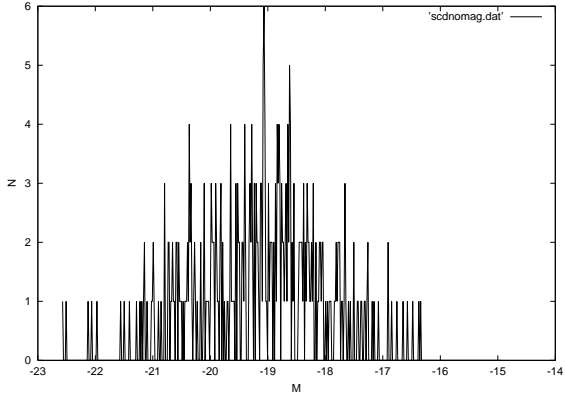}
\caption{The same as fig.~7 for about 339 Scd type galaxies.}
\end{figure}

\begin{figure}
\vbox to2.5in{\rule{0pt}{2.5in}}
\includegraphics{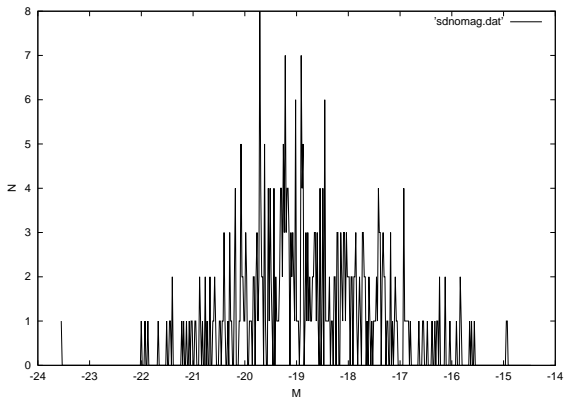}
\caption{The same as fig.~7 for the Sd type galaxies.}
\end{figure}

\begin{figure}
\vbox to2.5in{\rule{0pt}{2.5in}}
\includegraphics{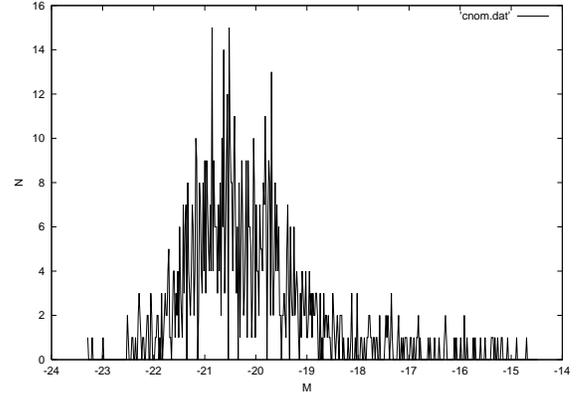}
\caption{The same as fig.~7 for about 450 C type galaxies.}
\end{figure}

\begin{figure}
\vbox to2.5in{\rule{0pt}{2.5in}}
\includegraphics{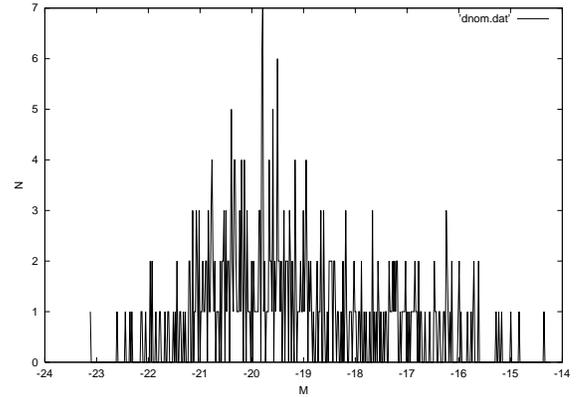}
\caption{The same as fig.~7 for about 230 D type galaxies.}
 \end{figure}

\newpage

\begin{table*}
 \centering
 \begin{minipage}{140mm}
\caption{In this table, the morphological type, SAM in units of $pc^2yr^{-1}$, the most
probable absolute magnitude ($M_{most}$), the average mass ($\bar{m}$) in units of
$10^{11}m_{\odot}$ and the most probable luminosity-average mass ratio in units of
${L_{\odot}\over M_{\odot}}$ are shown.}
\begin{center} 
\begin{tabular}{|l|l|l|l|l|}
\hline
${L_{most\over\bar{m}}} $& $\bar{m}$& $M_{most}$&$\bar{SAM} $&$morphs$ \\
\hline
0.425 &0.46 &-20.99&0.32&Sab \\
\hline
0.391 &0.44&-20.85&0.405&Sbc\\
\hline
0.416 &0.41 &-20.84 &0.428 &Sc \\
\hline
0.121 &1.22 &-20.68 & 0.448 &Sa \\
\hline
0.491 & 0.37 &-20.91&0.468&Sb  \\
\hline
0.429 &0.12  &-19.54 &0.494 &Scd  \\
\hline
0.471 &0.08 &-19.21 &0.504&Sd    \\
\hline
 \end{tabular}
 \end{center} 
\end{minipage}
\end{table*}

\section{CONCLUSIONS}

We have used the LEDA data base with complete morphological classification to study
the phenomenological investigation in the framework of the GFC model.
We conclude as follows:
\\
a) Within the same morphological type the expected behavior of SAM in terms of luminosity
is not seen. However, this is not surprising because of difference in masses of galaxies.
Further, it is shown that when we use SL instead of L for different masses of galaxies,
the discrepancy is removed by compensating the role of mass. Therefore, we receive
confirmation for GFC model.
\\
b) For different morphologies, SFR of each type has a significant  role in the
value of the most probable luminosity per unit average mass. This may not be looked
as a discrepancy with GFC model, because the SFR history depends on SAM showing
a weak form of the "decay mechanism" inherent in the GFC model, for galaxies, too.
\\
c)The distribution of compact galaxies in terms of their absolute magnitude,
shows higher average luminosities compared with diffuse galaxies. This is
another aspect of the GFC model.

\section*{Acknowledgments}
We are grateful to professors Y. Sobouti, A. Kembahvi and G. Swarup  for their
helpful comments. This work is supported by Institute for Advanced Studies in Basic
Sciences, IASBS, Zanjan, Iran.

\bsp

\label{lastpage}

\end{document}